# Nonlinear reflection from the surface of neutron stars and puzzles of radio emission from the pulsar in the Crab nebula

V. M. Kontorovich

Institute of Radio Astronomy, National Academy of Sciences of Ukraine, 4 Art Str., Kharkov 61002, Ukraine and V. N. Karazin Kharkov National University, 4 Svobody Square, Kharkov 61022, Ukraine
E-mail: vkont@rian.kharkov.ua

## Annotation

Having no any explanations the radiation of high-frequency components of the pulsar in the Crab Nebula can be a manifestation of instability in the nonlinear reflection from the neutron star surface. Reflected radiation it is the radiation of relativistic positrons flying from the magnetosphere to the star and accelerated by the electric field of the polar gap. The discussed instability it is a stimulated scattering by surface waves, predicted more than forty years ago and still nowhere and by no one had been observed.

## 1. Introduction

Neutron stars were predicted by Landau,[1] related to supernovae by Baade and Zwicky,[2] and discovered a quarter century later in the form of pulsars by Hewish et al.[3] Produced by collapse in supernova explosions, these stars have very high magnetic fields of $10^{12}$ G, rotate rapidly (with periods ranging from seconds to milliseconds), and are blanketed in a magnetosphere of electron-positron pairs which mostly corotates with the star but contains a beam of open lines of force over the magnetic poles, along which particles are accelerated and emit electromagnetic radiation.[4] Particles are accelerated in the gap above the region of the open lines of force,[1] which contains a strong accelerating electric field generated by the magnetic field and the rotation. This polar cap region is bounded by the lines of force of the magnetic field tangent to a luminous cylinder where the corotation velocity approaches the speed of light. Neutron stars have a nuclear density in which neutron production reactions $p+e^- \to n+\nu$ take place and contain superfluid and, possibly, superconducting layers.[5] The properties of matter at these nuclear densities have not known sufficiently well, so there is a number of different theoretical models.[6] Very little is known about the properties of the surface. It appears to be close to ideally conducting or, on the contrary, to be a good dielectric. In some pulsars it has a solid crust that is subject to starquakes. Given the tremendous gravita-tional force, the surface is nearly specular, but may contain a regular structure of elevations owing to the influence of the strong magnetic (and electric) field. In the polar cap region the surface may be perturbed significantly by the incident radiation. In this region, the upper layer, which is heated by accelerated particles and radiation, may be in a liquid state and lie on top of the solid crust (see p. 110 of Ref. [4]).

[1] An outer gap (not discussed in this model) between regions of the magnetosphere with electrical charges of different signs also plays an important role.

The physical idea behind the present paper will evidently yield additional information on the surface of a star based on its reflective properties. The idea of reflected radiation from the surface[7] is the basis of the proposed mechanism for the radio emission from the pulsar in the Crab nebula. The radiation from relativistic positrons flying toward the star from the magnetosphere is reflected. The reflected radiation predominates in the frequency range, where the interpulse shift and the high-frequency components HFC1 and HFC2 are observed.[8] The interpulse shift is considered to be an argument for the good reflective properties of that surface.

The reverse motion of the positrons[9] takes place in the accelerating electric field of the gap.[10,11] It has been examined before in connection with heating of the star's surface. At present most researchers are inclined to assume that the radio emission originates in the depth of the magnetosphere or beyond its limits, near the "luminous cylinder."[12,13] We, on the other hand, are interested in the radiation emerging from the inner gap above the polar cap, for which there is sufficient justification.[14] The mechanism for the radiation changes as the frequency varies.[15] In particular, as we assume, within a certain range of radio frequencies the emission from positrons flying into the star,[7] observed in the form of "light spots" (emission of the shifted interpulse and high-frequency components HFC1 and HFC2) upon reflection from the star surface, becomes predominant.

## 2. Frequency changes in the spectrum of the Crab pulsar

The pulsed emissions from the pulsar in the Crab nebula, for which unique multi-frequency measurements are available,[8,16] consists of a main pulse and an interpulse, as well as additional pulsed components at several frequencies. One of the mysteries of these data is the discovery by Moffett and Hankins[8] of a shift in the position of the interpulse at high radio frequencies (centimeter band) compared to the lower frequencies and a return to its previous position at still higher optical and x-ray frequencies.

An explanation for these frequency changes in the interpulse reflection has been proposed by Trofimenko and the author[7] in terms of reflection of the emission from relativistic positrons that may predominate over direct electron emission, thereby leading to a change in the interpulse emission mechanism.

The need to include the reflection of emission from the positrons is dictated by the observed frequency changes in the intensity and position of pulses from the Crab pulsar, which has not been physically explained up to now. These changes show up in the figures in Moffett and Hankins' paper[8] (Fig. 1), where a gradual "disappearance" of the lowfrequency interpulse and the main pulse can be seen, along with a shift in the high-frequency interpulse and the appearance of two high-frequency pulsed components.

### 3. Change in the emission mechanisms

An explanation for the "disappearance" of the main pulse has been proposed by Kontorovich and Flanchik.[15] It involves shutting down of the "nonrelativistic" emission mechanism with a broad directional diagram during longitudinal acceleration of electrons in the gap. As the electrons reach relativistic velocities, this mechanism becomes weaker and is shut off at sufficiently high frequencies (see circles on Fig.1.).

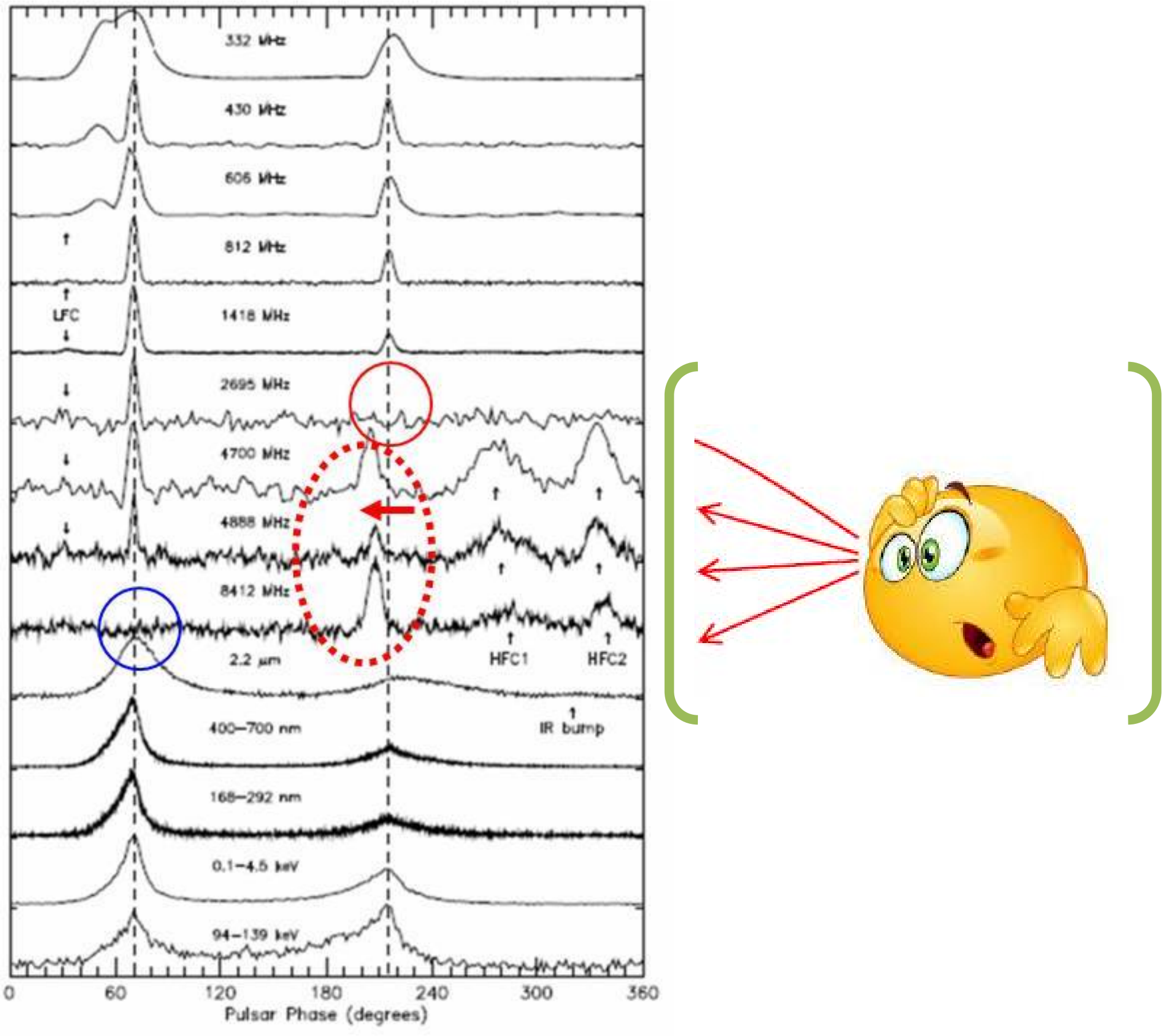

**FIG. 1**. Multifrequency observations of PSR 0531+21 (data of Moffett and Hankins,[8] with thanks to the authors). Dotted ellipse shows the shift of IP position. The thin arrows point on the HFCs

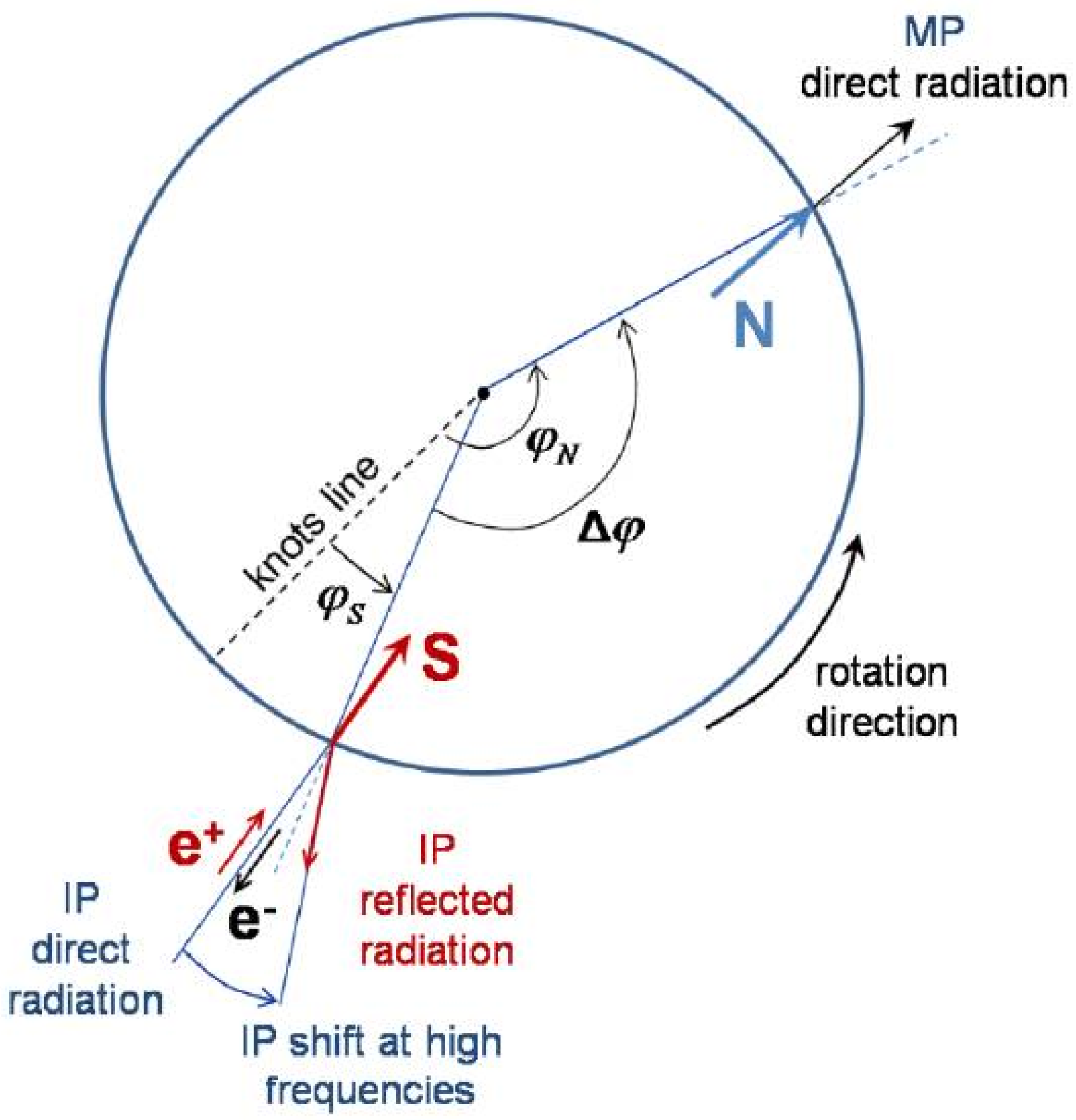

**FIG. 2.** Illustrating the shift in the interpulse (IP) during reflection at the pole S with an inclined magnetic field.[7] For angles of incidence (reflection) and for a downward shift of the interpulse, in the following we use the notation $\theta$ in order to avoid introducing more subscripts on angles $\varphi$. From the paper [7] by the courtesy of my coauthor S. V. Trofymenko

The change in mechanism corresponds to a break in the emission spectrum discovered by Malofeev and Malov[17,18] in an analysis of a catalog of pulsar spectra. At frequencies above the break, the well-known relativistic curvature radiation with a narrow directional diagram comes into play[13]. In this way, geometrical factors associated with the topology of the magnetic field become important.

### 3.1. The shift in the interpulse and the appearance of high-frequency components as a manifestation of the reflection of positron radiation

In a crude qualitative model, we shall assume that the source is the radiation from relativistic positrons moving along the magnetic field toward the surface.[7] In order to explain the shift of interpulse within the same frequency range, where the high-frequency components appear, it is required that the S pole of the magnetic field corresponding to the interpulse be inclined at an angle equal to half the phase shift of the interpulse, $\theta_S \approx 7^\circ$ (Fig. 2). This means that the wave is incident on the surface close to the direction of the normal, at an angle $\theta_S/2$. The specularly reflected wave should be deflected by the same angle in the direction opposite to the direction of rotation, which leads to a shift in the position of the interpulse relative to that of the electron emission. This determines the direction of the inclination of the magnetic field in the pole (Fig. 2). The appearance of the high-frequency component HFC1 (Fig. 1), as we shall assume below, is related to Raman scattering of the radiation on the surface, with the angle of inclination determined by the inclination of the magnetic field,[2] i.e., half the angular displacement of the interpulse, $\theta/2$.

When the incident wave interacts with surface waves, an antistokes wave appears in the first order perturbation theory; it is the result of the merging of these waves with the combined (summarized) tangential wave vector. Under the conditions of interest here, this wave has to be grazing (the Wood anomaly, see below). The surface wave is amplified, creating a periodic relief on the star's surface. Scattering on the periodic surface creates a Stokes reflected wave with the difference tangential wave vector. Under these conditions, this Stokes wave propagates, unlike the grazing antistokes wave. It will be counter propagating (Fig. 3) and deviating by a large angle from the normal (and from the direction of the magnetic field) along the direction of rotation. It is easy to find this angle (neglecting refraction), and we shall do this below. We shall assume that this wave is responsible for the appearance of the highfrequency component HFC1 (see Section 4). We may relate (with a less confidence) the component HFC2 to a second pole N that is responsible for the main pulse (MP), which is not seen in this range. In our model, therefore, the shift of the interpulse and the appearance of the high-frequency components are interrelated. They are determined by reflection on the surface of emission from relativistic positrons incident on it. Here, the shift in the interpulse is a specular reflection, while the high-frequency components are produced by a nonlinear effect, so called 'stimulated scattering', which is discussed further in the following.

It can, therefore, be hoped a study of the high-frequency components together with the shift in the interpulse will yield further information on the properties of neutron stars (pulsars) in the region of their polar caps. This interpretation can also be an argument in favor of the idea that the radio emission from pulsars in certain frequency bands develops in an inner (polar) gap.

**4. Stimulated scattering on the surfaces of neutron stars**

In order to discuss the physical idea upon which this article is based, we examine reflection from a periodically modulated surface, initially neglecting the anisotropy of the dielectric properties of the medium induced by the magnetic field.

---

[2] The relativistic positrons move strictly along the magnetic field and radiate within a narrow cone in the direction of their motion.

### 4.1. Combination fields: The action of surface waves

It is known from the classical papers of Mandel'shtam[19] and Andronov and Leontovich[20] that when light is scattered from the interface between two media with densities $\rho^{I.II}$ and dielectric constants $\varepsilon^{I.II}$, besides the mirror reflection components $\mathbf{E}_0^{R,T} \propto \exp(i\mathbf{k}_0^{R,T}\mathbf{r} - i\omega_0 t)$ when a surface deformation wave is present the fields of the form $\mathbf{E}_{\pm m}^{R,T} \propto \exp(i\mathbf{k}_{\pm m}^{R,T}\mathbf{r} - i\omega_{\pm m} t)$, develop at the combination frequencies $\omega_{\pm m} = \omega_0 \pm m\,\mathrm{Re}\,\Omega(q)$, $m = 1,2...$, with the tangential components of the wave vectors satisfying the condition $\mathbf{k}_{\pm m\,t}^{R,T} = \mathbf{k}_{0t} \pm m\mathbf{q}$, where $\Omega$ and $\mathbf{q}$ – are the frequency and wave vector of a surface wave (SW), e.g., gravitational, capillary, or Rayleigh, while $\omega_0$ and $\mathbf{k}_0$ are the same for the incident wave, and the indices $R$ and $T$ denote the reflected and refracted waves, respectively. In the following we only consider the diffraction components with numbers $\pm 1$ and assume that $\omega_0 >> \Omega$, $\mu = 1$. The scattered field are found from the boundary conditions at the surface $z = \zeta(x,y,t) = \zeta_{q\Omega}\exp(i\mathbf{q}\mathbf{r} - i\Omega t)$, where $\zeta_{q\Omega}$ — are the amplitudes of the SW,

$$\left[\mathbf{n}, \mathbf{E}^I - \mathbf{E}^{II}\right]_{z=\zeta(x,y,t)} = 0, \quad \left[\mathbf{n}, \mathbf{H}^I - \mathbf{H}^{II}\right]_{z=\zeta(x,y,t)} = 0 \qquad (4.1)$$

Here $\mathbf{n}$ – is the normal to the surface $z = \zeta(x,y,t)$. Taking $k_0|\zeta| << 1$ and $q|\zeta| << 1$, in a linear approximation in $\zeta$ we find out from (1) the fields $E_{-1} \propto \zeta^* E_0$, $E_1 \propto \zeta E_0^*$ and so on, which are bilinear in the amplitudes of the SW and incident field:[21,22]

$$\mathbf{E}_{-1}^{R,T} = -(1/2)i\zeta(\varepsilon - 1)\left[\mathbf{C}_{-1}^{R,T}E_{0y}^T + \mathbf{B}_{-1}^{R,T}H_{oy}^T\right], \qquad (4.2)$$

$$\mathbf{E}_{1}^{R,T} = -(1/2)i\zeta(\varepsilon - 1)\left[\mathbf{C}_{1}^{R,T}E_{0y}^T + \mathbf{B}_{1}^{R,T}H_{oy}^T\right],$$

where

$$C_{-1x}^{R,T} = -a_{-1}k_{-1x}k_{-1y}, \quad C_{-1y}^{R,T} = a_{-1}(k_{-1x}^2 - k_{-1z}^R k_{-1z}^T), \quad C_{-1z}^{R,T} = a_{-1}k_{-1y}k_{-1z}^{T,R},$$

$$B_{-1x}^R = d_{-1}\left[\varepsilon k_{0x}k_{-1x}k_{-1z}^R - k_{0z}^T(k_{-1z}^R k_{-1z}^T - k_{-1y}^2)\right],$$

$$B_{-1y}^R = d_{-1}k_{-1y}(\varepsilon k_{0x}k_{-1z}^R - k_{-1x}k_{-1z}^T),$$

$$B_{-1z}^R = d_{-1}\left[\varepsilon k_{0x}(k_{-1x}^2 + k_{-1y}^2) - k_{-1x}k_{0z}^T k_{-1z}^T)\right],$$

$$\varepsilon = \varepsilon^{II}/\varepsilon^I, \quad a_{-1} = (k_{-1z}^T - \varepsilon k_{-1z}^R)^{-1}, \quad d_{-1} = a_{-1}c/\omega\varepsilon^{II}.$$

Expressions for $\mathbf{B}_{-1}^T$ are obtained from $\mathbf{B}_{-1}^R$ on making the substitution $k_{-1z}^T \leftarrow k_{-1z}^R$ and dropping the multiplier $\varepsilon$ in front of $k_{0x}$. Formulas for $\mathbf{C}_1^{R,T}$ и $\mathbf{B}_1^{R,T}$ are obtained from those for $\mathbf{C}_{-1}^{R,T}$ and $\mathbf{B}_{-1}^{R,T}$ on replacing the index –1 by 1.

Later on, we shall be interested in the case of large module $\varepsilon$. In the coefficients for the combination fields, this modulus shows up in the numerator as the "nonlinear element" ($\varepsilon - 1$) and in the denominator (the coefficient $a_{\pm}$), in the form of a factor with $k^R_{\pm 1z}$. Thus, for $\varepsilon >> 1$ the factors compensate one another in the estimate for the combination fields if only $k^R_{\pm 1z}$ is not small. For the grazing components, the amplitudes are becoming much greater (the Wood anomaly). In fact, $a_{-1} = (k^T_{-1z} - \varepsilon k^R_{-1z})^{-1}$ for $k^R_{-1z} = 0$ transforms to $1/k^T_{-1z} \approx 1/k\sqrt{\varepsilon}$. That is, instead of $\varepsilon$, the denominator now contains $\sqrt{\varepsilon}$. Thus, while the estimate for the combination fields far from the Wood anomaly is of the form $E_{\pm} \approx (k\varsigma)E_{0y}$, where $k$ is the wave number of the incident wave, the anomaly for the grazing wave gives $E_{\pm} \approx \sqrt{\varepsilon} \cdot (k\varsigma)E_{0y}$; i.e., the grazing combination fields increase by a factor of $\sqrt{\varepsilon} >> 1$. Thus, later we shall examine only the case corresponding to this anomaly, limiting ourselves for definiteness to the case of transparent medium.

### 4.2. Nonlinear effects. Reverse influence of fields on surface waves

The effect being examined here involves the reverse influence both of the scattered and incident fields on the motion of the boundary. It can be significant for high incident wave intensities. The 'radiation pressure' wave that develops at the boundary is bilinear in the amplitude of the incident and scattered waves,

$$p_{cв} \propto E_0 E^*_{-1} + E^*_0 E_1 \propto \exp(i\mathbf{qr} - i\omega t), \tag{4.3}$$

and, in turns, drives oscillations of the surface.[21,22]

For a liquid medium, we solve the linearized equation of motion for an incompressible fluid $\rho \partial \mathbf{v} / \partial t = -grad\, p + \rho \mathbf{g}$ including forces acting on the part of electromagnetic fields[23,24] with the following boundary conditions at $z = \zeta$:

$$p^{II} - p^{I} - \alpha(\partial^2/\partial x^2 + \partial^2/\partial y^2)\zeta = p_{cв}, \quad p_{cв} \equiv \Pi^I_{nn} - \Pi^{II}_{nn}.$$

Here $p \equiv p' - E^2(8\pi)^{-1}\rho\partial\varepsilon/\partial\rho$, $\Pi_{nn} -$ is the normal component of the Maxwell stress tensor; $p'$, $\mathbf{v}$, and $\alpha$ — are the pressure, velocity, and surface tension coefficient; and $\boldsymbol{g}$ is the acceleration of gravity. This yields the Fourier component of the deflection $\zeta$, which is expressed in terms of the Fourier component of the radiation pressure,

$$\zeta_{\mathbf{q}\Omega} = |q|(p_{cв})_{\mathbf{q}\Omega} / (\rho^I + \rho^{II})\left[\Omega_0^2(q) - \Omega^2\right], \tag{4.4}$$

where $\Omega_0(q)$ is the unperturbed dispersion relation for the surface[3] waves. The amplitude of the pressure wave at frequency $\Omega$ is given by

$$p_{cв} = iq\zeta P\varepsilon^I \left|E_0^i\right|^2 / 8\pi \qquad (4.5)$$

with

$$\begin{aligned} P = \frac{(\varepsilon-1)}{4q}\Big\{ & T_s^2 \cos^2\varphi \cdot (C_{1y}^T - C_{-1y}^{T*}) + \\ & +\varepsilon^I T_p^2 \sin^2\varphi \left[ Z_x (B_{1x}^T - B_{-1x}^{T*}) + \varepsilon Z_z (B_{1x}^T - B_{-1x}^{T*}) + 2q_x(\varepsilon-1) Z_x Z_z \right] - \\ & -\sqrt{\varepsilon^I}\, \frac{T_s T_p}{2} \left[ B_{1y}^T - B_{-1y}^{T*} + Z_x (C_{1x}^T - C_{-1x}^{T*}) + \varepsilon Z_z (C_{1z}^T - C_{-1z}^{T*}) + 2q(\varepsilon-1) Z_z \right] \sin 2\varphi \Big\} \end{aligned} \qquad (4.6)$$

where $T_s = E_{0y}^T / E_{0y}^i$, $T_p = H_{0y}^T / H_{0y}^i$ – are the Fresnel coefficients, $\varphi$ – is the angle between the vector $\mathbf{E}_0^i$ of the incident wave and the y axis, and $Z_{x,z} = E_{0x,z} / H_{0y}$, while the index $i$ denotes the amplitude of the incident wave. The dispersion relation for the SW on the irradiated surface is found using Eqs. (3) and (4) and including the damping owing to the (low) viscosity $\nu = \eta / \rho^{II}$ (Refs. [3] and [23])

$$\Omega(q) = \pm\Omega_0(q) - 2iq^2\nu \mp \frac{iq^2 P\varepsilon^I \left|E_0^i\right|^2}{16\pi\rho^{II}\Omega_0(q)} \qquad (\rho^{II} >> \rho^I). \qquad (4.7)$$

For an incident wave of intensity exceeding the threshold

$$\varepsilon^I \frac{\left|E_0^i\right|^2}{8\pi} > \frac{4\eta\,\Omega_0(q)}{\left|\mathrm{Re}\,P\right|} \qquad (4.8)$$

surface waves are pumped and stimulated Raman scattering takes place on them. An analysis of the threshold reduces to studying the quantity $\mathrm{Re}\,P$ (5) and (6), which is proportional to the radiation pressure. An examination of stimulated Raman scattering on Rayleigh SW in an isotropic solid (with small opto-elastic constants) yields a dispersion relation (7) in which $\Omega_0(q) = c_r q$, where $c_r$ – is the Rayleigh wave velocity,[25] and a dispersion term of the same order of magnitude as for bulk acoustic waves. If it is written in the form $\eta_{eff} q^2$, then the estimated threshold for stimulated Raman scattering outside the resonance (as before) has the form $E_0^2 / 8\pi \approx \eta_{eff}\Omega_0(q)$.

---

[3] In this example, they are capillary-gravitational waves. Other waves are quite possible on the magnetized surface of the polar cap, such as surface MHD waves, spin waves, etc.

## 5. High-frequency Moffett-Hankins components as a result of stimulated scattering

In this interpretation, a surface wave **q** leading to the Wood anomaly[26,27] is predominantly excited by positrons incident on the surface; during scattering on it, the first order $k^+$ antistokes component becomes glancing (we have simplified the notation, omitting several indices). For convenience, we limit ourselves initially to the case with a SW wave vector lying in the plane of incidence, so that intuitive diagrams can be used (see Fig. 3).

An angle of incidence close to the angle of inclination of the magnetic field at the pole[7] is the Rayleigh angle in this case. The reciprocal influence of the Stokes and antistokes Raman scattering components corresponding to the maximum growth rate of the stimulated scattering pumps forward and reverse 'Wood waves'. This leads to growth in the antistokes wave, as well as of the Stokes component reflected from the surface, which propagates at a certain angle of "reflection" $\theta_S^{-R}$ at the S pole. We also assume that this wave produces the high-frequency component in the emission of the pulsar B0531+21 observed by Moffett and Hankins. Thus, in this model, the phase lag in the component HFC1 is associated with the displacement of the interpulse and should be observed in the same frequency range.

We now calculate this phase[4] beginning with the known displacement $\theta_S$ of the interpulse. According to a paper[7] on the relation between the reflection of the interpulse and specular reflection of positron radiation in the S pole, the angle of incidence in this model is given by $\theta_S/2$. Denoting the wave vector of the Wood wave by $\boldsymbol{q}$ and limiting ourselves to scattering in the plane of incidence, from $k_t^{+R} = k$, $k_z^{+R} = 0$ we obtain

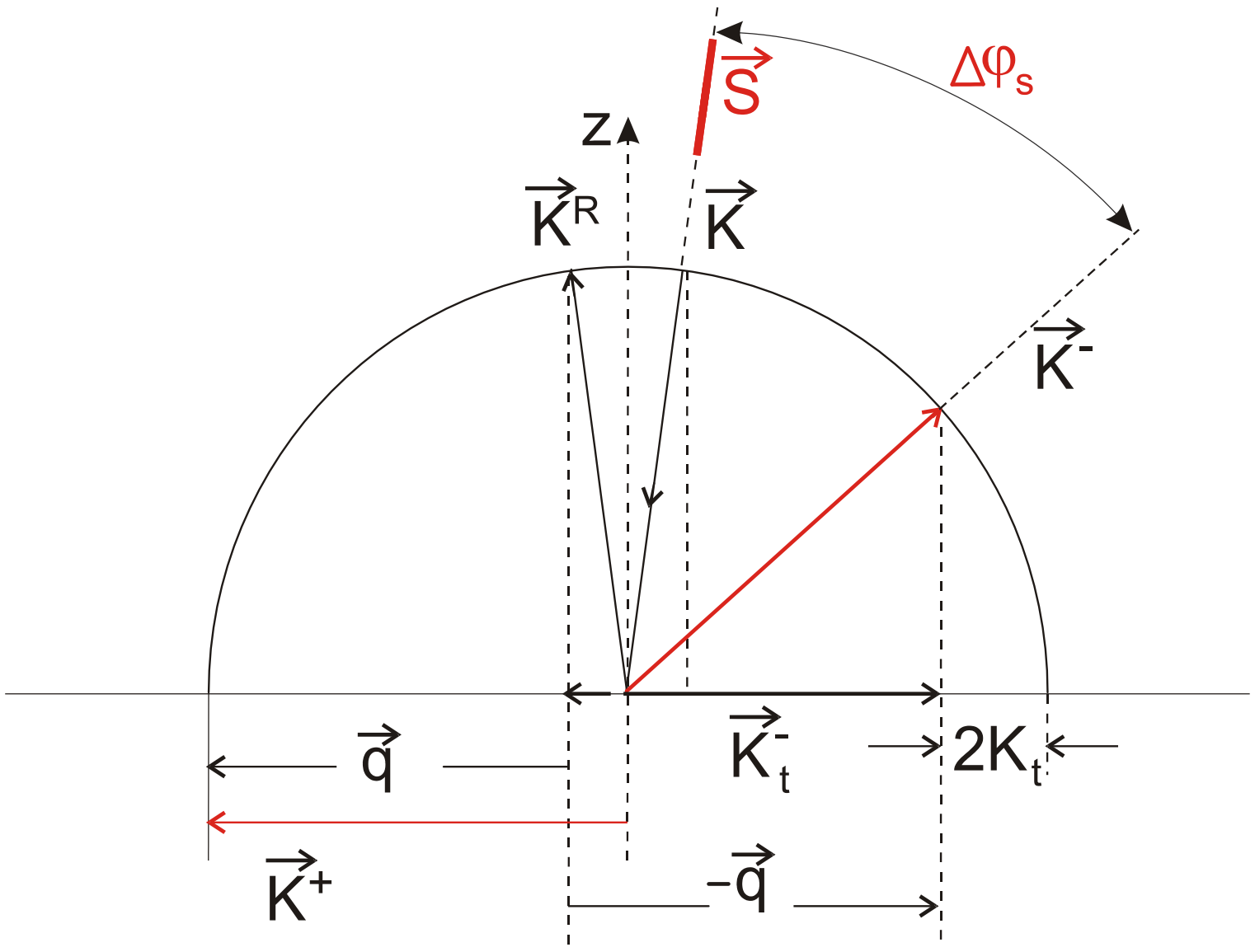

**FIG. 3.** Illustrating the appearance of first order Raman spectra in the pole S. The direction of the incident wave is determined by the direction of the magnetic field. The wave vector of the 'Wood wave' surface q: when the incident wave merges with it, a grazing antistokes wave $\mathbf{k}^+$ (the Wood anomaly) is produced. The amplified surface wave creates a propagating Stokes wave $\mathbf{k}_-$ which is presumably observed as the high-frequency component HFC1.

---

4 We use the angle $\varphi$ to indicate coordinates of the component and $\theta$ for the displacement of the interpulse and angles of incidence.

$$\mathbf{q} = \mathbf{k}^{+} - \mathbf{k_t}\,, \tag{5.1}$$

where $k_t = k\sin\frac{\theta_S}{2}$ is the tangential component of the wave vector of the incident radiation, determined by the angle of inclination of the magnetic field. The latter is found from the shift in the interpulse. For the Stokes component of interest to us,

$$k_t^{-R} = k - 2k_t = k\cdot(1 - 2\sin\frac{\theta_S}{2})\,. \tag{5.2}$$

Thus, the "angle of reflection" $\theta^{-R}$ for it follows from the condition: (see Fig.3)

$$\sin\theta_S^{-R} = \frac{k - 2k_t}{k} = 1 - 2\sin\frac{\theta_S}{2}\,. \tag{5.3 }$$

We have used the condition that the frequency of the Wood wave is low compared to the incident wave frequency. When this frequency does make a contribution, the equations become

$$\omega^{\pm} = \omega(\mathbf{k}) \pm \Omega(\mathbf{q}),\ \mathbf{k}_t^{\pm} = \mathbf{k}_t \pm \mathbf{q}\,. \tag{5.4 }$$

We take the shift of the interpulse to be $\theta_S = 7^\circ$.[8] According to Eq. (11), this gives a phase shift of $\Delta\varphi_S = 69^\circ$ for the HFC1. HFC1 is shifted to a frequency of 8.9 GHz and $\Delta\varphi^{HFC1} \approx 79^\circ$.[8] The discrepancy is $\Delta\varphi^{HFC1} - \Delta\varphi_S \approx 10^\circ$. This could already be considered a good agreement, given the crudity of the model and the $\approx 30^\circ$ width of the high frequency component.

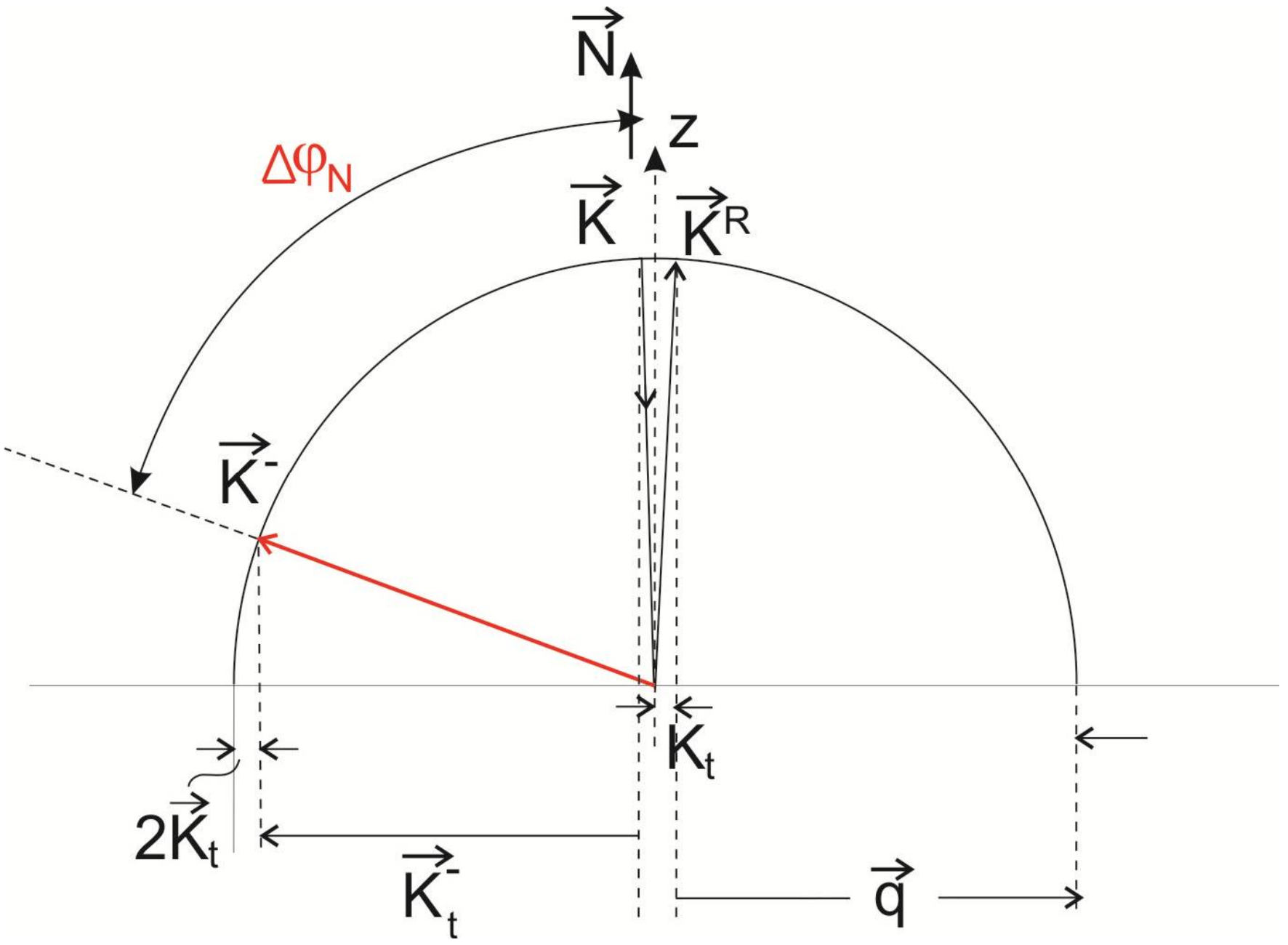

**FIG. 4**. Illustrating the possible appearance of first order Raman spectra at the **N** pole. In order to distinguish the detailed better, we have chosen a large angle of incidence.

It is natural to try to relate the second high-frequency component to the N pole, which corresponds to the main pulse (Fig. 4). Since the main pulse emission does not enter the telescope within this frequency interval (the pulse "disappears") and its (expected) displacement is unknown to us, we can recover its parameters relying on data for HFC2. The phase shift for this component relative to the reverse side from the N pole is close to $90^\circ$:[8] $\Delta\varphi^{HFC2} \approx 91^\circ$. We introduce the same difference in angles (e.g., by refraction) as for the interpulse: $\Delta\varphi^{HFC2} - \Delta\varphi_N \approx 10^\circ$. Then $\Delta\varphi_N \approx 81^\circ$. Using this value, we find the (invisible to us) displacement $\theta_N$,

$$\theta_N = 1 - \sin\Delta\varphi_N \approx 1 - 0.988 = 0.012 = 0.5^\circ \qquad (5.5)$$

This, of course, is a very rough estimate which assumes the same value of the calculated and observed phase shifts of the component as at **S**, as well as identical physical conditions at both poles and on both paths from the reflection (scattering) point to the point at which it emerges from the magnetosphere of the pulsar.

## 5.1. Possible effect of a resonance with a surface electromagnetic wave

We shall use this classical terminology,[24] but the term 'surface polariton' is also used in the literature.[28] The surface properties of a neutron star and, especially, the properties of the thin surface layer of the polar cap in pulsars (with a thickness of a few centimeters, so that the effects discussed here take place at centimeter wavelengths) of interest to us are not known in detail (see the discussion and references in Ref. 4, p. 110), but we can assume that a surface electromagnetic H-wave can exist on its surface[24] and it can come into resonance with a grazing wave. In our analysis, this corresponds to the denominators of the combination reflection coefficients going to zero, i.e., the quantities $a_l$ in Eq. (13)

$$k_{1z}^T = \varepsilon k_{1z}^R, \qquad (5.6)$$

which because of the different signs of $k_{1z}^T$ and $k_{1z}^R$, is possible for $\varepsilon = \varepsilon_I / \varepsilon_{II} < 0$; i.e., for different signs of the dielectric constants on the two sides of the interface. The position of the Wood resonance differs from that of the Rayleigh point corresponding to a simple Wood anomaly (the appearance of a grazing component) and the maximum growth rate is shifted. An attempt can be made to relate this displacement to the magnitude of the deviation $\Delta\varphi^{HFC1} - \Delta\varphi_N$.

The dispersion relation ($k_{sf}$ is the wave number of the surface electromagnetic wave) derived from Eq. (14) is, of course, the same as that obtained for the homogeneous problem[5]

$$k_{sf}^2 = k_0^2 \frac{\varepsilon^I |\varepsilon^{II}|}{(|\varepsilon^{II}| - \varepsilon^I)}, \quad k_0 = \frac{\omega}{c}, \qquad (5.7)$$

where the conditions $\varepsilon^I \cdot \varepsilon^{II} < 0$ and $|\varepsilon^{II}| > \varepsilon^I$: must be satisfied. Setting $\varepsilon^I = 1$ and $|\varepsilon^{II}| >> 1$, we arrive at $k_{sf}^2 \approx k_0^2 (1 + |\xi|^2)$, where $\xi = \sqrt{1/\varepsilon^{II}}$ is the surface impedance. The resonance Wood wave, therefore, corresponds to an additional shift $\Delta q \approx k_0 \cdot |\xi|^2 / 2$. For large inclinations of the surface $q\zeta$ compared to the impedance, the former influence the magnitude and position of the resonance.[29,30] This sort of shift (also in the presence of dissipation) has been observed in experiments[31,32] on the reflection of electromagnetic waves by a diffraction grating under the conditions of the Wood anomaly.[6]

The Wood resonance can, therefore, in principle explain the observed discrepancy in the position of the phase of the high-frequency component. This is not, however, the only reason for the shift. Thus, the shift obtained here only provides an upper bound estimate of $|\xi|^2 \le 0.1$: for the surface impedance. Normal refraction should give a shift in the same direction (since the charge density of the Goldreich-Julian magnetosphere plasma falls off with height). An HFC beam propagates in the plane where the angle between the axes of rotation and the magnetic field is close to $90^{\circ}$, so that the plasma density is low. Preliminary estimates of the shift owing to refraction give an acceptable magnitude of the shift. This estimate must, of course, be improved with a more realistic model of the refractive index[33] and magnetosphere.

---

[5] The same result is obtained for the homogeneous problem,[24] which corresponds to an approach to zero by the denominator in the expressions for the combination fields (2).

[6] In individual cases, such as an almost ideal conducting medium, the resonance character of the grazing scattered wave leads to such a large increase in its amplitude that it becomes necessary to consider the essentially nonlinear problem drawing on higher order diffraction spectra than the first. The relationships among the various small parameters[29,30] become important and affect the shift in the resonance.

## 6. Conclusion

This has been an examination of the simplest isotropic non-dissipative model for the dielectric properties of the medium and for scattering in the plane of incidence. These limits must be avoided in the case of pulsars, and multiwave behavior in the pulsar magnetosphere must be taken into account. The possibility that the HFC2 is related to this effect must be considered. The broad spectrum of the highfrequency components may be indicative of a significant excess beyond the instability threshold and the excitation of a large number of waves, interactions among which can be significant. Thus, we may be dealing with wave turbulence. The effect of refraction requires detailed study.

The proposed treatment of the high-frequency Moffett- Hankins components in the Crab pulsar is an argument in favor of emission mechanisms in the polar gap, in particular with longitudinal acceleration of electrons[15] which can predominate within a certain range of frequencies. The explanation for interpulse shift and for the highfrequency components in the emission from the pulsar B0531+21 in terms of reflection of (positron) radiation from the pulsar surface proposed in [7] and here will hopefully open up additional possibilities for the study of neutron stars. In addition, an interpretation of the Moffett-Hankins components in terms of stimulated scattering on surface waves would both mean that the effect was observed (in exotic circumstances for nonlinear optics) and demonstrate that stimulated scattering takes place in nature. If such forms of nonlinear scattering as volume stimulated Brillouin scattering, etc., led to powerful observed energy transfer between Raman modes, then nonlinear scattering on surface waves should show up in connection with the surface structures created by irradiation. We note that the redistribution of energy during reflection of waves from the surface of a diffraction grating which forms the basis of the explanation of the Wood anomaly has only been measured quantitatively and clearly demonstrated only very recently.[31,32] The above discussion raises a number of questions about nonlinear diffraction of some relevance to the problem discussed here: in particular, the redistribution of the reflected energy fluxes over various channels during excitation of direct and reverse Wood waves and, therefore, the combined effect of the Stokes and antistokes waves.

I thank my coauthors, colleagues, and friends V. K. Gavrikov, I. S. Spevak, and A. V. Kats, as well as S. V. Trofimenko, E. Yu. Bannikova, and M. V. Nikipelov, for discussing various aspects of this paper and also for great help in setting the text, drawing the figures, and giving permission to use parts of the texts of our joint papers.

## Appendix: Stimulated scattering and surface structures

Stimulated scattering, which was discovered shortly after the invention of lasers, is as general a phenomenon as the nonlinear shift in the frequency of an oscillator in a strong external wave field. Stimulated scattering is essentially an instability manifested by almost all lowfrequency modes in a sufficiently powerful (laser) field, i.e., as established long ago, every spontaneous scattering event has a corresponding stimulated analog. The growth rate characterizing stimulated scattering is nothing other than the positive imaginary part of a frequency shift corresponding to an influx of energy into a given degree of freedom and its loss by nonlinear damping. The stimulated scattering growth rate is proportional to the incident field intensity.

The stimulated analog of spontaneous scattering on surface waves was first discussed independently by groups from Kharkov and Moscow.[34,35] It was shown that the maximum growth rate corresponds to the appearance of one or more grazing scattered waves.[21] At approximately the same time, it was found experimentally that during laser irradiation of metal, semiconductors, and dielectrics accompanied by significant thermal effects, periodic structures (i.e., a lattice [36-38]) are formed on their surfaces within a certain range of pulse intensities and durations.

The period of these lattices depends on the angle of incidence and is on the order of magnitude of the wavelength of the applied light, while the orientation of the periodic structure is determined by the polarization direction and angle of incidence of the light. The periodic structures are formed by scattering of an electromagnetic wave on seed irregularities at the surface of the material with subsequent interference of the incident and scattered light and a reverse effect of the interference pattern on the slow movements of the medium (boundary), just as happens in stimulated scattering, but because of thermal mechanisms (in nondissipative media the pump mechanism arises from radiation pressure). It is remarkable that the parameters of the periodic structures are not coupled in such detail with the specific mechanism of the dissipative interaction, but depend mainly on the scattering geometry, which determines the structure of the heat source found by examining the electrodynamics part of the problem.

In the text used the translation of D. H. McNeill with author's additions.